\documentclass{aa}  

\usepackage{graphicx}
\usepackage[dvipsnames]{xcolor}
\usepackage{multirow}

\usepackage{txfonts}
\usepackage[]{hyperref}
\hypersetup{
    colorlinks=true,
    citecolor=blue,
    linkcolor=blue,
    urlcolor=blue,
    }
\newcommand{\orcid}[1]{\href{https://orcid.org/#1}{\includegraphics[width=10pt]{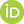}}}

\begin{document}

   \title{Progenitor mass and ejecta asymmetry of SN 2023ixf from nebular spectroscopy}

   \author{Lucía Ferrari
          \inst{1,2}
          \thanks{luciaferrari@fcaglp.unlp.edu.ar}
          \orcid{0009-0000-6303-4169}
          \and
          Gastón Folatelli\inst{1,2,3}\orcid{0000-0001-5247-1486}
          \and 
          Keila Ertini\inst{1,2}\orcid{0000-0001-7251-8368}
          \and
          Hanindyo Kuncarayakti \inst{4,5}\orcid{0000-0002-1132-1366}
          \and
          Jennifer E. Andrews \inst{6}\orcid{0000-0003-0123-0062}
          }

   \institute{Instituto de Astrofísica de La Plata, Paseo del Bosque S/N, 1900, Buenos Aires, Argentina
         \and
             Facultad de Ciencias Astronómicas y Geofísicas Universidad Nacional de La Plata, Paseo del Bosque S/N B1900FWA, La Plata, Argentina
        \and
            Kavli Institute for the Physics and Mathematics of the Universe (WPI), The University of Tokyo, Kashiwa, 277-8583 Chiba, Japan
        \and
            Finnish Centre for Astronomy with ESO (FINCA), 20014 University of Turku, Finland
        \and
            Tuorla Observatory, Department of Physics and Astronomy, 20014 University of Turku, Finland
        \and
            Gemini Observatory/NSF's NOIRLab, 670 N. A'ohoku Place, Hilo, HI 96720, USA
             }

   \date{Received April 18, 2024; accepted Month XX, XXXX}

 
  \abstract
   {SN 2023ixf was discovered in galaxy M101 on May 2023. Its proximity made it an extremely valuable opportunity for the scientific community to study the characteristics of the SN and its progenitor. 
   A point source detected on archival images and hydrodynamical modelling of the bolometric light curve has been used to constrain the former star's properties. 
   There is a significant variation in the published results regarding the initial mass of the progenitor.
   Nebular spectroscopy provides an independent tool to enhance our understanding of the supernova and its progenitor.}
   {We determine the SN progenitor mass by studying the first published nebular spectrum taken 259 days after the explosion. }
   {We analyze the nebular spectrum taken with GMOS at the Gemini North Telescope. Typical emission lines are identified, such as [\ion{O}{i}], H$\alpha$, [\ion{Ca}{ii}], among others. Some species' line profiles show broad and narrow components, indicating two ejecta velocities and an asymmetric ejecta. We infer the progenitor mass of SN 2023ixf by comparing with synthetic spectra and by measuring the forbidden oxygen doublet flux.}
   {Based on the flux ratio and the direct comparison with spectra models, the progenitor star of SN 2023ixf had a $M_{ZAMS}$ between 12 and 15 $M_\odot$.
   In comparison with this, we find the use of the [\ion{O}{i}] doublet flux to be less constraining of the progenitor mass.
   Our results agree with those from hydrodynamical modelling of the early light curve and pre-explosion image estimates pointing to a relatively low-mass progenitor.}
   {}

   \keywords{supernovae:general --- 
supernovae: individual: SN 2023ixf}

   \maketitle
%

\section{Introduction}\label{sec:intro}

One key aspect of astrophysics is understanding the fate of massive stars and the origin of supernova (SN) explosions. Hydrogen-rich SNe, known as Type II SNe (SNe~II), arise from the core collapse of stars more massive than about 8 $M_\odot$ during the red supergiant (RSG) phase, as has been confirmed through direct progenitor detections {\citep{maund2005,vandyk2012,smartt2015}}. However, there is an apparent contradiction between theory and observations on the high-end range of SN-II progenitor masses. Direct detections appear to show a lack of progenitors more massive than $17-18$ $M_\odot$ while the theoretical limit should be around $25-30$ $M_\odot$ {\citep{smartt2015}}. Although several explanations have been proposed to explain this apparent controversy {\citep{davies2018}}, it is of utmost importance to determine the masses of SNe-II progenitors.

Apart from direct detections on pre-explosions images, which is a powerful method albeit limited to the nearby Universe, there are other more indirect ways to derive progenitor properties. One such method is through the modelling of the nebular-phase spectra. Once the plateau phase ends, most of the SN ejecta become transparent and the spectral emission originates from regions near the former stellar core. Analysis of the nebular spectrum reveals properties of the core, such as its mass {\citep[][]{Jerkstrand2012,Kuncarayakti2015,Dessart2021}}. The stellar-core mass is in turn indicative of the initial mass with which the star was formed. Furthermore, the shape of the emission lines provides information about the distribution of the innermost ejecta and thus can give hints on the possible asymmetries that may occur during the explosion {\citep{taubenberger2009,fang2022}}.

Type II SN~2023ixf, discovered in M101, was one of the most nearby SNe discovered in the last few decades. Due to its proximity, it raised the interest of professional and amateur astronomers alike. A veritable wealth of articles and communications have been issued since its discovery in May 2023. Part of these analyses focused on early-time time observations across the electromagnetic spectrum. The presence of a dense circumstellar material surrounding the SN was revealed by early-on high ionization "flash" spectral lines \citep{perley2023,jacobsongalan2023,smith2023,bostroem2023,hiramatsu2023,teja2023}, along with an early excess in the light curve, especially in the ultraviolet range {\citep{jacobsongalan2023,hosseinzadeh2023,teja2023,martinez2023}}, and X-ray, radio and polarimetry observations {\citep{grefenstette2023,berger2023,vasylyev2023}}. Several groups analyzed pre-explosion observations the SN site from the Hubble Space Telescope and the Large Binocular Telescope in the optical, and the Spitzer Space Telescope in the infrared. In all, these works find a putative progenitor object at the SN location whose spectral energy distribution is compatible with it being a dust-enshrouded RSG star. However, estimates of the zero-age main sequence (ZAMS) progenitor mass differed widely, from $M_\mathrm{ZAMS} \approx 8$ $M_\odot$ to $M_\mathrm{ZAMS} \approx 18$ $M_\odot$ 
(\citealt[][$18.1^{+0.7}_{-1.2}~M_\odot$]{qin2023};
\citealt[][$12-15~M_\odot$]{vandyk2023};
\citealt[][$11\pm2~M_\odot$]{kilpatrick2023};
\citealt[][$17\pm4~M_\odot$]{jencson2023};
\citealt[][$12^{+2}_{-1}~M_\odot$]{xiang2023};
\citealt[][$9-14~M_\odot$]{neustadt2024}). Additional analyses of the stellar populations in the vicinity of the SN provided estimates of $M_\mathrm{ZAMS}=17-19$ $M_{\odot}$ \citep{niu2023}, and $M_\mathrm{ZAMS} \approx 22$ $M_{\odot}$ \citep{liu2023}. A progenitor mass of $M_\mathrm{ZAMS} = 20 \pm 4$ $M_{\odot}$ was found by \citet{soraisam2023} by interpreting the observed variability of the pre-SN source in the IR observations as due to pulsational instability of an M-type star.
Independently of the pre-explosion observations, \citet{bersten2024} performed a hydrodynamical modelling of the bolometric light curve and photospheric velocity evolution and were able to constrain the progenitor mass to $M_\mathrm{ZAMS} < 15$ $M_\odot$. 

In this work, we present the first nebular spectrum of SN~2023ixf as an additional attempt to constrain the progenitor mass using independent information. The spectrum also serves to provide some insights on the internal structure of the SN ejecta through line profile analysis. Section~\ref{sec:observations} provides a description of the observational data. 
In Section~\ref{sec:spec} we analyze the nebular spectrum properties and we estimate a progenitor mass in Section~\ref{sec:prog_mass}. Our conclusions are summarized in Section~\ref{sec:conclusion}.

\section{Observations, reddening and distance} \label{sec:observations}
The spectrum presented in this work was obtained with the Gemini Multi-Object Spectrograph \citep[GMOS;][]{hook2004} mounted on the Gemini North Telescope (Program GN-2024A-Q-139, PI Andrews).
The observations were divided into four $900-$s exposures in long-slit mode with the R400 grating. The data were processed with standard procedures using the IRAF Gemini package. Flux calibration was performed with a baseline standard observation.
The spectrum resolution is $\approx 10~\AA$ ($\approx 500$ km s$^{-1}$ at $6000~\AA$). 
Adopting the explosion time at $\mathrm{JD} = 2460083.25$ \citep{hosseinzadeh2023}, the spectrum phase is 259 days after the explosion.


The redshift published in the NASA/IPAC Extragalactic Database\footnote{\url{https://ned.ipac.caltech.edu/}} (NED) for the host galaxy and used in this work is 0.0008. We additionally adopt a distance to M101 of $6.85 \pm 0.15$ Mpc \citep{Riess2022}, and a Milky-Way reddening of $E(B-V)_\mathrm{MW} = 0.008$ mag \citep{Schlafly2011}. Reddening in the host galaxy was estimated by \citet{Lundquist2023} to be $E(B-V)_\mathrm{Host} = 0.031$ mag. We applied the extinction correction by using a standard extinction law \citep{Cardelli1989} with $R_V = 3.1$ and thus the total $E(B-V)_\mathrm{Total} = 0.039$ mag.

\section{Spectral properties} \label{sec:spec}

\begin{figure*}
    \centering
    \includegraphics[width=1\linewidth]{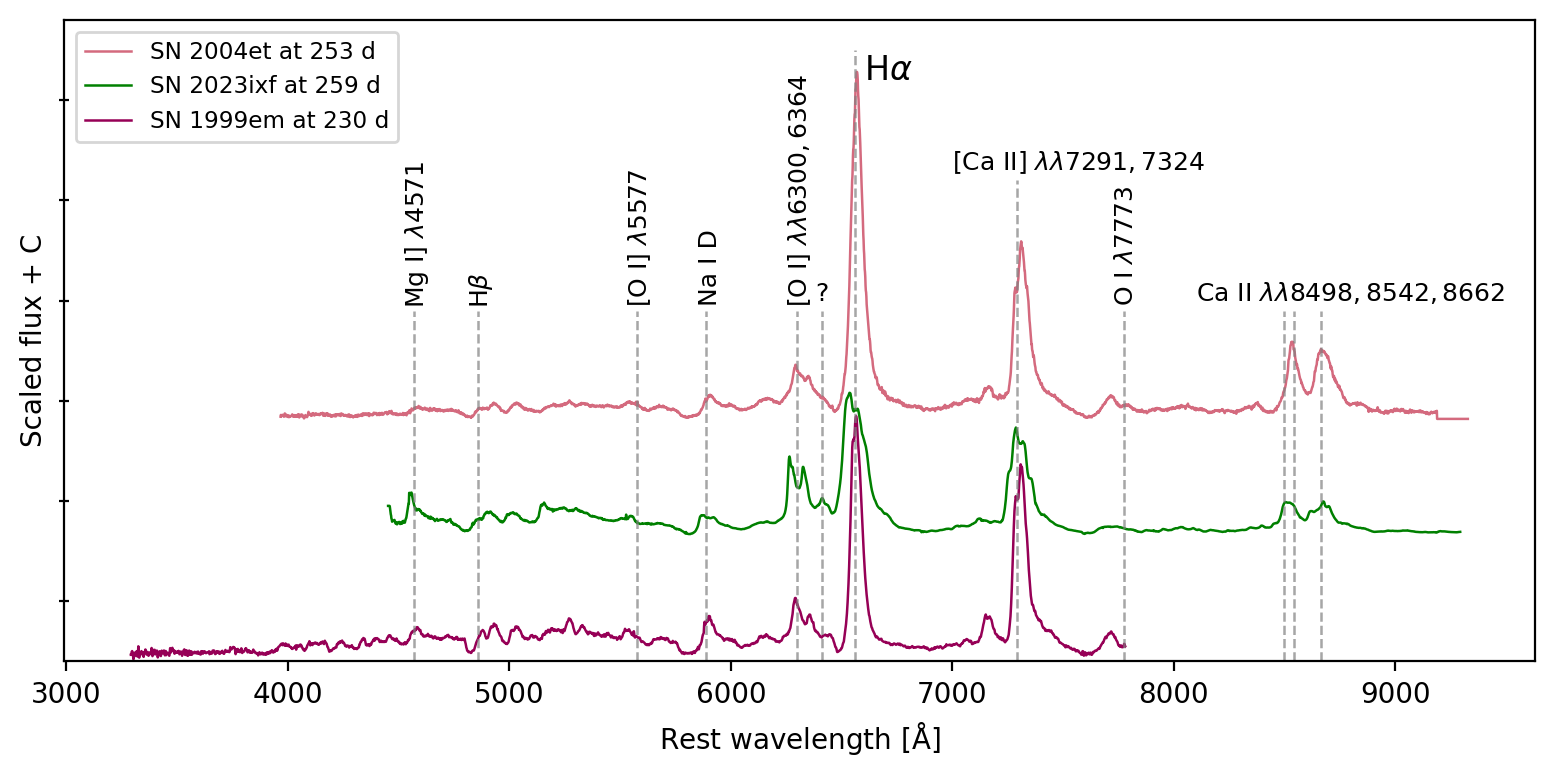}
    \caption{Nebular spectra of SN~2023ixf compared with those of SN~1999em \citep{leonard2002} and SN~2004et \citep{Sahu2006} at similar phases. The spectra are corrected for redshift and not extinction. Prominent emission features are indicated with gray vertical lines.}
    \label{fig:line_ID}
\end{figure*}

The nebular spectrum of SN~2023ixf is presented in Figure~\ref{fig:line_ID}, in comparison with two prototypical Type IIP SNe, SN 1999em at 230 d \citep{leonard2002} and SN~2004et at $253$ d \citep{Sahu2006}.
Typical spectral features of SNe II at this phase are seen in the spectrum, most prominently the emissions from H$\alpha$, [\ion{O}{i}]$\lambda\lambda6300,6364$, and [\ion{Ca}{ii}]$\lambda\lambda7291,7324$. Also identifiable are emission lines due to \ion{Mg}{i}]$\lambda$4571 and the Ca IR triplet (\ion{Ca}{ii}$\lambda\lambda$8498,8542,8662), and the P-Cygni profile due to the \ion{Na}{i}~D doublet (\ion{Na}{i}$\lambda\lambda$5890,5895).
No emission line is identifiable as H$\beta$. The [\ion{O}{i}] feature appears stronger relative to H$\alpha$ and the [\ion{Ca}{ii}] doublet as compared with SN~1999em and SN~2004et (see more on this in Section~\ref{sec:prog_mass}).
Except for the [\ion{Ca}{ii}] doublet, the most prominent lines appear more blueshifted in SN~2023ixf than in its counterparts.

After comparing with the sample of Type II-P SNe from \citet{Silverman2017}, we find that the emission lines appear in general broader in SN~2023ixf than in the other objects, except for SN 1992ad and SN 1993K. 
This is especially true in the case of H$\alpha$, which in SN 2023ixf shows a complex profile.
 More on this will be discussed in Sections~\ref{sec:line_profiles} and \ref{sec:geometry}.

\subsection{Line profiles}
\label{sec:line_profiles}

\begin{figure}
    \centering
    \includegraphics[width=1\linewidth]{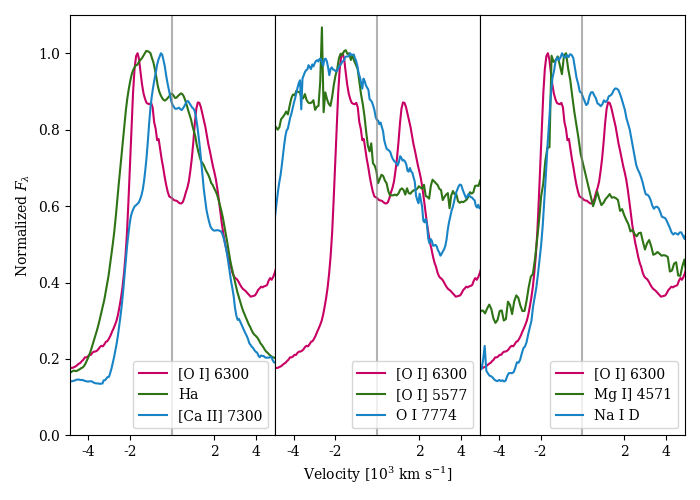}
    \caption{Line profiles of several emission lines in 2023ixf.}
    \label{fig:profiles}
\end{figure}

As shown in Figure~\ref{fig:line_ID}, most emission lines appear blueshifted in SN~2023ixf, except for [Ca~II]\,$\lambda\lambda7291,7324$ that is centered nearly at rest.
The line profiles of the main spectral features are shown in detail in Figure~\ref{fig:profiles}. In the case of H$\alpha$ the profile can be decomposed into two components, a wide one (with $\mathrm{FWHM} = 5300$ km~s$^{-1}$) centered at rest, plus an intermediate-width one ($\mathrm{FWHM} = 1600$ km~s$^{-1}$) shifted by $-1800$ km~s$^{-1}$.

\begin{figure}
    \centering
    \includegraphics[width=1\linewidth]{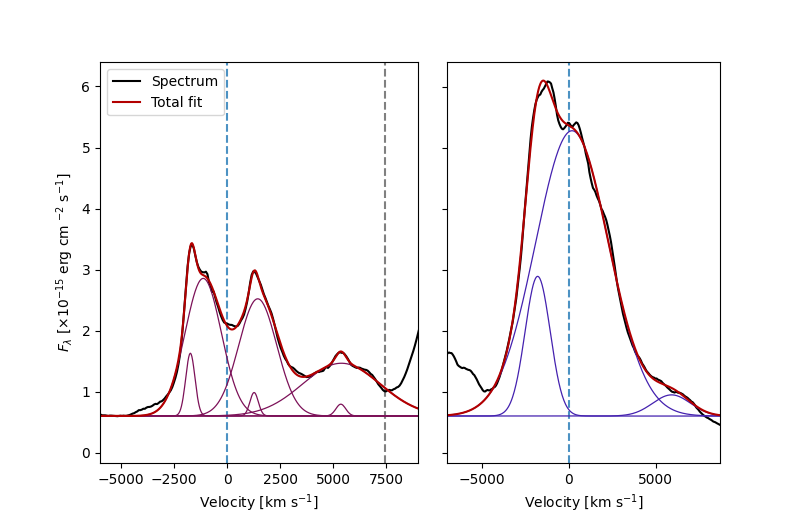}
    \caption{Fitting of the [O I] doublet and H$\alpha$ profiles. Blue dashed lines indicate the rest wavelength for 6300 \AA~and 6564 \AA~in the left and right panels respectively, and the grey dashed line indicates the rest wavelength for 6456\AA. The thin lines show the individual gaussian components.}
    \label{fig:O_fit}
\end{figure}

The [O~I]\,$\lambda\lambda6300,6364$ feature exhibits the double-peak profile that is typically seen in nebular spectra of core-collapse SNe \citep{Maeda2008,Modjaz2008,taubenberger2009}. The peaks are displaced roughly symmetrically relative to 6300 \AA.
Both emission peaks appear to have complex profiles that can be decomposed into a wider ($\mathrm{FWHM} \approx 2000$ km~s$^{-1}$) and a narrower ($\mathrm{FWHM} \approx 500$ km~s$^{-1}$\footnote{The width of these narrower lines is near the spectral resolution achieved by the data, so it may be considered as an upper limit to the actual width.}) component, as shown in Figure~\ref{fig:O_fit}. The centers of the wider components are separated by $\approx 2500$ km~s$^{-1}$, at $-1100$ km~s$^{-1}$ and $1400$ km~s$^{-1}$, respectively. 
The narrower peaks are slightly further apart with a separation of $\approx\,3000$ km~s$^{-1}$, and both peaks shifted by $-1700$ km~s$^{-1}$ and $1300$ km~s$^{-1}$, respectively. We note that such a separation of $\approx\,3000$ km~s$^{-1}$ is roughly equal to that between the [O~I]\,$\lambda6300$ and [O~I]\,$\lambda6364$ lines that produce this feature. 
Similar blue-shifts in [\ion{O}{i}] and H$\alpha$ were observed in SN 2007it \citep{Andrews2011}.
Furthermore, a multiple-component feature in [\ion{O}{i}] has been seen before in SN 2016gkg, and has been attributed to multiple distinct kinematic components of material at low and high velocities \citep{Kuncarayakti2020}.
Hypotheses on the material distribution in SN 2023ixf are discussed in Section \ref{sec:geometry}.

In order to fit the four components that we attribute to [O~I]\,$\lambda\lambda6300,6364$, we subtracted an additional set of wide and narrow Gaussian profiles centered at 6413 \AA\ (see Figure~\ref{fig:O_fit}, left panel). This feature may also be present in the spectrum of SN~1999em (see Figure~\ref{fig:line_ID}), and it may be attributed to O~I\,$\lambda 6456$ blueshifted by $\approx 2000$ km~s$^{-1}$. However, this identification is not clear as the profile may be affected by a remainder absorption component of the H$\alpha$ line to the red. There is no clear explanation that could account for this permitted line. Furthermore, we do not recognize this feature in the sample of nebular spectra from \citet{Silverman2017}.

Two emission features appear that may be identified as [\ion{O}{i}]$\lambda5577$ and \ion{O}{i}$\lambda7774$, although with a blueshift of $1500$ km s$^{-1}$. This is shown in Figure \ref{fig:profiles}, central panel. If the identifications are correct, the blueshifts are similar to those of the [\ion{O}{i}]$\lambda\lambda63006364$ lines.
\ion{Mg}{i}]$\lambda4571$ and \ion{Na}{I} D are clearly identifiable in the spectrum, both blue-shifted by $\approx 1000$ km~s$^{-1}$, roughly similar to the other lines. \ion{Na}{I} D shows a broad profile and an absorption produced by the interstellar medium (ISM) on top.

\subsection{Ejecta geometry}\label{sec:geometry}

The [\ion{O}{i}] double-peaked profile described above is an indicator of an asymmetry in the ejecta \citep{Kuncarayakti2020,fang2022}. It is also noticeable that the [\ion{Ca}{ii}] doublet is centered at rest wavelength. Thus, the disposition of the inner material could match that proposed by \citet{Fang2024}, in which an axisymmetric oxygen-rich torus surrounds a bipolar calcium region (see their Figure 3).
The general blueshift common to all species supports this hypothesis, meaning that we may be mostly seeing the torus's near side while the far side is mostly obscured. 
In addition to this, an outer, roughly spherical, high-velocity H-rich material must be responsible for the wide, rest-centered H$\alpha$ component.
Nevertheless, the two global, intrinsically different emitting regions of the ejecta, associated with the low- and high-velocities, may be challenging to explain in the context of the mechanism of the explosion, which takes place deep in the core. Here we highlight that the double-component (broad + narrow) line profile possibly occurs in all ejecta layers, from the outermost parts (H) down to the inner layers (O, Na, Mg) and core (Ca).

Another hypothesis that may explain these peculiar line profiles is the presence of an asymmetric or clumpy CSM that slows down parts of the ejecta and thus produces narrow, blue-shifted line components. Moreover, the presence of CSM surrounding the SN has already been proposed for SN 2023ixf \citep[e.g.][]{bostroem2023,jacobsongalan2023,Martinez2024}. The extended, H-rich envelope is mainly unaffected and seen as a rest-centered profile, while a small part of the hydrogen and the regions rich in other elements are mostly affected by the CSM slow-down.

\section{Progenitor mass}
\label{sec:prog_mass}

Nebular SN spectra are useful indicators of the progenitor mass. The flux of the [\ion{O}{i}]$\lambda\lambda6300,6364$ feature has been proposed as an indicator of CO core mass by proxy of the O mass in the core \citep{Uomoto1986,Jerkstrand2014}. Given that absolute fluxes can be affected by errors in the spectral calibration or in the distance and extinction estimations, the flux ratios of [O~I]\,$\lambda\lambda6300,6364$ over [Ca~II]\,$\lambda\lambda7291,7324$ have been proposed as more robust indicators of the progenitor mass \citep[][]{Fransson1989,fang2022}. The relation between the [O~I]/[Ca~II] flux ratio and the progenitor ZAMS mass is seen in the model nebular spectra by \citet{Jerkstrand2014}. Here we will derive a progenitor mass estimate by means of line fluxes and flux ratios, and by direct comparison with available model spectra.

\begin{figure}
    \centering
    \includegraphics[width=1.\linewidth]{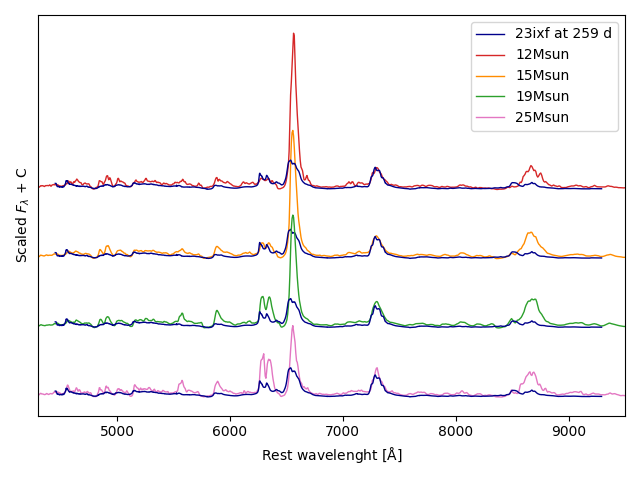}
    \caption{SN 2023ixf at 259 d compared to nebular spectra models from \citet{Jerkstrand2014} at 250 d.}
    \label{fig:Jerkstrand14}
\end{figure}

In Figure~\ref{fig:Jerkstrand14} we show the 259~d spectrum of SN~2023ixf and modeled spectra computed by \citet{Jerkstrand2014} for SNe~II at a similar age and for varying progenitor masses. The model spectra were scaled to match the distance to SN~2023ixf. A scaling factor of $1.15$ was applied to the observed spectrum in order to match the synthetic VRI magnitudes with the observed ones available from the American Association of Variable Star Observers (AAVSO\footnote{\url{https://www.aavso.org}}) International Database. The value was calculated by taking an average of the scaling factor for each filter. From the Figure, it can be seen that the observed spectrum has suppressed emission of permitted lines from H$\alpha$ and Ca~II~IR triplet as compared with all of the modeled spectra. Nevertheless, if one compares the emissions due to [O~I] and [Ca~II], the modeled spectra from 12 and 15 $M_\odot$ progenitors provide the closest matches to the observations.

In terms of the [O~I]/[Ca~II] flux ratio the value obtained from the observed spectrum is $0.51$. Compared with ratios of $0.35$, $0.60$, $1.18$, and $1.74$ obtained from the modeled spectra with 12, 15, 19, and 25 $M_\odot$, respectively, the result is indicative of a ZAMS progenitor mass between 12 and 15 $M_\odot$. Furthermore, high-mass progenitors with $M_\mathrm{ZAMS} \gtrsim 19$ $M_\odot$ are disfavored based on this comparison.

{We also considered the nebular spectra models from \citet{Dessart2021} to double-check this result. Although the epochs differ in about 100 days, as the available models are calculated for 360 days post-explosion, the oxygen-to-calcium ratio does not show significant time variation in the spectra models from \citet{Jerkstrand2014}. The measured ratio value for the SN 2023ixf is also consistent with progenitors with ZAMS mass below 15 $M_\odot$ compared with measurements performed on the models from \citet{Dessart2021}. The direct comparison also suggests a relatively low-mass progenitor, but we do not show it here as the nebular spectrum evolves significantly within 100 days. Further observations will enable us to use these models with more accuracy.}

Another way of estimating the progenitor mass comes from the direct measurement of the [\ion{O}{i}] doublet flux which constrains the minimum oxygen mass responsible for such emission. This mass is linked to the progenitor mass by theoretical oxygen production yields. This procedure is detailed in \cite{Jerkstrand2014} and has been used as an independent way of estimating the progenitor mass \citep[e.g.][]{Sahu2011,Kuncarayakti2015}. As mentioned before, the main uncertainties of this method lie in the absolute flux calibration (affected by the reddening correction and photometry scaling).

Assuming the emission with peak at 5550 $\AA$ is [\ion{O}{i}]$\lambda5577$ blueshifted by $\approx 1500$ km s$^{-1}$, we can establish a lower limit for the flux, as the continuum is not clear and the line might be blended with the [Fe I]$\lambda 5501$ line.
Consequently, the lower limit for the material temperature is 3700 K \citep{Jerkstrand2014}. 
The oxygen doublet flux is $2.38 \times 10 ^{-13}$ erg s$^{-1}$ cm$^{-2}$, giving an oxygen mass of 0.5 $M_\odot$. We assume this is the total mass that produces the measured flux. According to the oxygen production yields published by \citet{Nomoto1997} and \citet{Limongi2003}, the progenitor must have been a star below 17 $M_\odot$. If we consider the production yields provided by \citet{Rauscher2002} and \citet{Sukhbold2016}, the progenitor mass is well below 15 $M_\odot$.
Because of the uncertainties involved in this last method (in the absolute flux calibration and the temperature determination, see \citealt{Jerkstrand2017}), we prefer the former approach based on direct model comparison and flux ratios to constrain the progenitor mass. We thus conclude that the progenitor must have been a star with ZAMS mass between 12 and 15 $M_\odot$.

\section{Conclusions}\label{sec:conclusion}

We present the first nebular spectrum of SN 2023ixf published in the literature, taken 259 days after the explosion. The observations were obtained with the Gemini North Telescope using GMOS.
The spectrum was useful to examine the inner regions of the ejecta through line profiles and to determine the progenitor mass following a procedure independent from those previously published for this SN.

The spectrum shows the typical emission lines seen in a Type II SN resulting from the elements hydrogen, oxygen, calcium, sodium, and magnesium. In comparison with other Type II SNe, SN 2023ixf presents a weaker H$\alpha$ emission, and no H$\beta$ is recognizable. The [\ion{O}{i}] doublet is blue-shifted and shows a double-peaked profile. This is a sign of a non-spherical ejecta and may be interpreted as a torus-shaped emitting region. By fitting the line we distinguish two wide components and two narrow components, also distinguishable as intermediate line peaks in the H$\alpha$ profile, and other lines as well. This feature, also detected in \ion{Mg}{i}]$\lambda4571$ and \ion{Na}{i} D, is attributed to a low-velocity ejecta component, caused either by two distinct ejecta regions or by parts of the material being partly slowed down by an asymmetric or clumpy CSM. H$\alpha$ wide component is rest-centered and must be produced by an outer, roughly spherical, H-rich material.

The main result of our work is the estimation of the progenitor mass from the nebular spectrum. Via spectral comparison with models by \citet{Jerkstrand2014} we found that the progenitor mass was between 12 and 15 $M_\odot$. This was confirmed by comparing the [\ion{O}{i}]/[\ion{Ca}{ii}] flux ratio between the spectrum of SN~2023ixf and those of the models. We also derived an upper limit of $15 - 17~M_\odot$ for the progenitor mass based on the [\ion{O}{i}] flux. However, this result is less constraining and less reliable than the comparison with model spectra. Our derived mass is in agreement with the estimates of \citet{bersten2024} from hydrodynamical modelling and also \citet{kilpatrick2023,xiang2023,vandyk2023} and \citet{neustadt2024} from pre-explosion images.

\begin{acknowledgements}
    We thank Dave Sand and Jeniveve Pearson, CoI's for the Gemini Program GN-2024A-Q-139, for sharing the observational data. 
    We thank the Gemini Observatory for encouraging fluid group interaction between Programs GN-2024A-Q-139 and GN-2024A-Q-309 teams.
    H.K. was funded by the Research Council of Finland projects 324504, 328898, and 353019.
    We acknowledge with thanks the variable star observations from the AAVSO International Database contributed by observers worldwide and used in this research. 
    Based on observations obtained at the international Gemini Observatory (GN-2024A-Q-139, PI: Andrews), a program of NSF's NOIRLab, which is managed by the Association of Universities for Research in Astronomy (AURA) under a cooperative agreement with the National Science Foundation. On behalf of the Gemini Observatory partnership: the National Science Foundation (United States), National Research Council (Canada), Agencia Nacional de Investigaci\'{o}n y Desarrollo (Chile), Ministerio de Ciencia, Tecnolog\'{i}a e Innovaci\'{o}n (Argentina), Minist\'{e}rio da Ci\^{e}ncia, Tecnologia, Inova\c{c}\~{o}es e Comunica\c{c}\~{o}es (Brazil), and Korea Astronomy and Space Science Institute (Republic of Korea). This work was enabled by observations made from the Gemini North telescope, located within the Maunakea Science Reserve and adjacent to the summit of Maunakea. We are grateful for the privilege of observing the Universe from a place that is unique in both its astronomical quality and its cultural significance.
\end{acknowledgements}

%
\bibliographystyle{aa} 
\bibliography{biblio} 
%

%


\end{document}